\documentclass{article}

\usepackage{amsmath}
\usepackage{amssymb}
\usepackage{amsthm}

\newcommand{\sd}{\mathrm{d}}

\usepackage{tensor}

\title{A structured analysis of Hubble tension}
\author{}

\theoremstyle{definition}
\newtheorem{assumption}{Assumption}

\begin{document}

\begin{center}
{\Large\sc A structured analysis of Hubble tension}\\[10ex] 
 \vspace*{1cm}
 Wim Beenakker$^{a,b}$,
 David Venhoek$^{a}$\\[1cm]
 {\it 
$^a$ {Theoretical High Energy Physics, Radboud University
Nijmegen, Heyendaalseweg 135, 6525~AJ Nijmegen, The Netherlands}\\
$^b$ {Institute of Physics, University of Amsterdam, Science Park 904, 1018 XE Amsterdam, The Netherlands}}\\
\end{center}

\begin{abstract}
As observations of the Hubble parameter from both early and late sources have improved, the tension between these has increased to be well above the 5$\sigma$ threshold. Given this, the need for an explanation of such a tension has grown. In this paper, we explore a set of 7 assumptions, and show that, in order to alleviate the Hubble tension, a model needs to break at least one of these 7, providing a quick and easy to apply check for new model proposals. We also use this framework to make a rough categorisation of current proposed models, and show the existence of at least one under-explored avenue of alleviating the Hubble tension.
\end{abstract}

\section{Introduction}

Over the last decade cosmological observations have significantly improved, both for early-time as well as late-time observations.
Early-time measurements by the Planck collaboration give a value of $H_0 = 67.36 \pm 0.54$ \cite{aghanimplanck}. This value is corroborated by the results published in \cite{DESDark} combining Dark Energy Survey (DES), Baryon Acoustic Oscillation (BAO) and Big Bang Nucleosynthesis (BBN) constraints to give a value of $H_0 = 67.4\pm 1.1$.

For late-time measurements, results from \cite{riesslarge,wongh0licow,freedmancarnegie,reidmegamaser, huanghubble} were combined in \cite{verdetensions} to give a combined estimate of $H_0 = 73.3 \pm 0.8$ (though with some caveats on the error bounds due to not fully accounting for shared contributions in the distance ladder). Furthermore, it is shown that these results, even when discarding some, still provide a $4-6\sigma$ tension between early and late-time values of $H_0$. There have since been further late-time measurements reported in \cite{shajibstrides,yuanconsistent}, but other than strengthening error bounds, these do not significantly change this picture.

This tension is large enough that it has created a need for an explanation. As such, there have been a myriad of models proposed that alleviate (part of) this tension, overviews of which are given in \cite{knoxhubble, divalentinocosmology}.

Currently, there are few general results placing formal conditions on what a model alleviating the Hubble tension should look like. The authors are only aware of \cite{jedamzikreducing}, providing a general result showing that a viable model needs to modify more than just the size of the sound horizon.\footnote{For the claim that a modification of the sound horizon is necessary, or variations on it, the authors are not aware of a formal argument showing all the assumptions neccessary for such a conclusion.} The goal of this paper is to provide a new general (model independent) result setting conditions on models alleviating the Hubble tension.

We do this by taking a higher level view of possible alleviating models. In section~\ref{sec:constraints}, we provide a set of 7 assumptions based on CMB observations and direct measurements of $H_0$ which combine to create a space of options in which the cosmological standard model $\Lambda$CDM is an extremum. This then allows us to show that our 7 assumptions exclude any model that has the potential to alleviate the Hubble tension, implying that any such model should break at least 1 of our 7 assumptions.

We then demonstrate one application of this result in section~\ref{sec:analyseexisting} by making a categorisation of the space of existing proposals by means of the assumptions they break. This provides us with two things: A somewhat reasonable categorisation of existing proposals, and an under-explored avenue for new solutions.

\section{Constraints on new theories}\label{sec:constraints}
We will start by making sufficient assumptions on our model of the universe and its equations of motion to be able to use the regular distance formula for the comoving angular diameter distance $D_A(z)$ of an object at redshift $z$.

Further assumptions are then needed to allow us to use the results from the Planck analysis of the Cosmic Microwave Background (CMB). Finally, we impose an at first glance reasonable restriction on the shape of the total matter density $\rho$, which will then result in limitations on the values for $H_0$ that are attainable.

\begin{assumption}\label{a:relativity}
The laws of general relativity, and in particular the Einstein Equations $R\indices{_\mu_\nu} - \frac{1}{2}Rg\indices{_\mu_\nu} = 8\pi G T\indices{_\mu_\nu}$, are a good approximation of the physics of the universe at cosmological scales.
\end{assumption}

Note that, with the above choice for the Einstein Equations, we absorb any cosmological constant $\Lambda$ into the stress-energy tensor $T\indices{_\mu_\nu}$.

\begin{assumption}\label{a:isotropy}
The universe can be approximated as spatially isotropic and homogeneous at large scales.
\end{assumption}

Assuming a spatially isotropic universe, behaving according to the laws of general relativity, the metric becomes the Friedmann-Robertson-Walker metric, and can be written in spherical coordinates as
\begin{align}
\sd s^2 &= -\sd t^2+a(t)^2\left[\frac{\sd r^2}{1-\kappa r^2}+r^2\sd\Omega^2\right].
\end{align}
Here, $a(t)$ is the scale factor of the universe at time $t$, and $\kappa\in\{-1,0,1\}$ indicates whether the metric is spatially hyperbolic, spatially flat, or spatially spherical respectively.

Given this form of metric, the Einstein equations reduce to the Friedmann Equation and covariant energy conservation equation, and can be written as
\begin{align}
H(t)^2 &= \frac{8\pi G}{3}\rho(t)-\frac{\kappa}{a(t)^2},\label{eq:friedmann1}\\
0 &= \partial_0\rho(t)+3H(t)(\rho(t)+p(t)),\label{eq:friedmann2}
\end{align}
with $p(t)$ the pressure, and $H(t) = \frac{\partial_0 a(t)}{a(t)}$ the Hubble parameter, giving as present-day value $H_0 \equiv H(0)$.

We choose to use the covariant energy conservation equation, instead of the second Friedmann Equation, in order to more easily reason about its impact on the behaviour of $\rho$ later. It is a relatively straightforward calculation to show that both forms are in fact equivalent in this situation.

\begin{assumption}\label{a:curvature}
The universe has no significant spatial curvature.
\end{assumption}

Ignoring the possibility of spatial curvature further simplifies the metric to
\begin{align}
\sd s^2 &= -\sd t^2+a(t)^2\left[\sd x^2 + \sd y^2 + \sd z^2\right].
\end{align}
and removes the curvature term from the first Friedmann Equation. This in turn implies that
\begin{align}
H &= \sqrt{\frac{8\pi G}{3}\rho}
\end{align}
in an expanding universe.

\begin{assumption}\label{a:redshift}
The relationship between photon redshift $z$ and scale factor $a$ is $a=\frac{1}{1+z}$.
\end{assumption}

This assumption now allows us to use $z$ as a time variable. Combined with the other assumptions made so far,
 it also fully constrains the relationship between $\rho(t)$, $D_A$ and $z_*$. This can be written as
\begin{align}
D_A &= \int_0^{z_{*}}\frac{\sd z'}{H(z')}\\
&= \sqrt{\frac{3}{8\pi G}}\int_0^{z_{*}}\frac{\sd z'}{\sqrt{\rho(z')}}\label{eq:distancerelation}
\end{align}

With this in place, we now turn to making the assumptions needed to bring in measured values for the comoving angular diameter distance $D_A$, the redshift of the CMB $z_*$, the matter density at decoupling $\rho_m(z_*)$ from Planck observations, and the directly measured value of $H_0$. By necessity, the assumptions leading up to these are somewhat broad:

\begin{assumption}\label{a:early}
The physics leading to the angular scale of the baryonic sound horizon, and its size, as well as the related estimates of total matter density at decoupling, are accurately approximated in the Planck analysis. In particular, the measured angular diameter distance to the CMB, the redshift of decoupling, and the matter density at decoupling, have no significant systematic or model-induced errors that are unaccounted for.
\end{assumption}

\begin{assumption}\label{a:late}
The physics related to distance measurements of late-time objects is well understood and provide an accurate way of measuring the Hubble parameter at late times. In particular, there are no significant systematic or model-induced errors on measured redshifts and absolute brightnesses unaccounted for.
\end{assumption}

Before we can start drawing conclusions, there is one final constraint needed on the behaviour of the total density. For this, we define the model specific density $\rho_x$ as
\begin{align}
\rho_x(z) = \rho(z) - \rho_m(z_*)\left(\frac{1+z}{1+z_*}\right)^3.
\end{align}

This definition splits off the non-interacting dilution of matter due to expansion from any other, model-specific, effects. Note that there is no specific meaning to $x$, rather we use $\rho_x$ generically to refer to the model-specific contributions to $\rho$. We then assume the following:

\begin{assumption}\label{a:growth}
$\frac{\partial \rho_x}{\partial z} \ge 0$ for $z < z_*$.
\end{assumption}

Intuitively, this assumption states that there is no net matter creation in $\rho_x$, and that there is no decay from $\rho_m$ to a component contributing to $\rho_x$ after the decoupling of the CMB.

Combining this final assumption into Equation~\ref{eq:distancerelation}, we find
\begin{align}
D_A &= \sqrt{\frac{3}{8\pi G}}\int_0^{z_{*}}\frac{\sd z'}{\sqrt{\rho_m(z_*)\left(\frac{1+z'}{1+z_*}\right)^3+\rho_x(z')}} \\
&\le \sqrt{\frac{3}{8\pi G}}\int_0^{z_{*}}\frac{\sd z'}{\sqrt{\rho_m(z_*)\left(\frac{1+z'}{1+z_*}\right)^3+\rho_x(0)}}.
\end{align}

As $H_0$ is directly related to $\rho(0)$, we find that the following holds under the assumptions made above: for a given value of $H_0$, the $\Lambda$CDM model has maximal comoving angular diameter distance to the CMB, and any alternative model results in a lower value for $D_A$. Since $D_A$ is fixed by measurements, this implies that for any model conforming to our assumptions with non-constant contribution $\rho_x$ to the total energy density of the universe, the value for \emph{$H_0$ would need to be lower} than that predicted by extrapolating from the $\Lambda$CDM model. As we find from observations that the currently measured value for \emph{$H_0$ is higher} than predicted from the $\Lambda$CDM assumptions in combination with the Planck observations, we can conclude that \emph{one of the assumptions 1-7 must be broken}.

\section{Broken conditions in existing theories}\label{sec:analyseexisting}

We can now reconsider existing models for alleviating the Hubble tension within the framework of our assumptions. This gives insight into how existing models achieve their alleviation of the Hubble tension, and provides insight into avenues of reducing the early-late time tension that are not yet as well explored.

In doing this, we will consider a variety of models from the literature in the coming pages. However, our focus here will be primarily on providing clarity in illustrating the various ways our assumptions can be broken, and as such, this will not be an exhaustive overview of theories alleviating the Hubble tension. The interested reader can find good starting points for further theories in~\cite{knoxhubble}.

Furthermore, in choosing these examples, we have not focused on whether such models are, or have the potential to be, consistent with further observational evidence beyond the constraints imposed by direct measurements of $H_0$ and measurements of the CMB. This choice was made specifically to provide a more complete illustration of the approaches already considered.

Finally, some of the examples given here could be classified under multiple of these assumptions, either because they break multiple conditions, or because one can reasonably disagree which of the conditions is actually broken. The classification here represents the opinions of the authors as to which condition(s) best reflect the way in which a model alleviates Hubble tension. Although others might classify these models differently, we believe that the resulting conclusions drawn from this classification are robust under these differing opinions.

\subsection{Systematic effects affecting measurements of $H_0$}
Let us first consider the option of systematics affecting late-time measurements, breaking assumption~\ref{a:late}. In the case of CMB observations by Planck, \cite{aghanimplanck}~indicates no direction for poorly or uncontrolled sources of systematic effects from known physics. For late-time measurements of $H_0$, the wide variety of independent observations (\cite{riesslarge,wongh0licow,freedmancarnegie,reidmegamaser, huanghubble}, since augmented with~\cite{shajibstrides,yuanconsistent}) also leads the authors of~\cite{verdetensions} to consider this to be an unlikely option.

However, there exist several effects that act similarly to late-time systematics within our framework of assumptions. This could be from effects such as photon brightening (section IV.2 in~\cite{knoxhubble}, originating from ideas in~\cite{csakidimming} and~\cite{meyerfermi}), it could be the result of a local matter void~\cite{dinggigaparsec}, or originate from screened fifth forces~\cite{desmondlocal}. These theories cause late-time measurements to overestimate $H_0$ by decreasing the apparent distance, reducing the tension directly that way. %TODO: Reevaluate change in final layout

\subsection{Modification of early-time physics}
At early times, there is similarly the option of systematics, or systematics-like influences of a new model, breaking Assumption~\ref{a:early}. In the case of CMB observations by Planck, \cite{aghanimplanck}~indicates no direction for poorly or uncontrolled sources of systematic effects from known physics. However, apparent systematics can result from effects changing the scale of the sound horizon, such as neutrino mass mechanisms~\cite{escuderocmb} or the effects of additional light particles~\cite{wymanneutrinos,aghanimplanck}. Such theories typically modify the time of matter-radiation equality, resulting in changes to the size of the sound horizon. So-called early dark energy models~\cite{karwaldark,poulinearly} achieve a change of the sound horizon by changing the evolution just after matter-radiation equality, giving similar net effects from the perspective of our reasoning. This changed sound horizon, which typically becomes smaller, in turn decreases the angular diameter distance, resulting in higher values of $H_0$ from early-time measurements.

These are not the only options considered in the literature. A second approach studied in~\cite{chianginferences, sekiguchiearly} involves modifications to the mechanism of recombination, resulting in a different temperature of recombination, or a more stretched recombination. These modifications change the interpretation of the CMB, resulting in a different angular distance and/or dark matter mass estimate.

\subsection{Subverting the spatially flat FRW framework}
A second approach to alleviating the Hubble tension is found in subverting the assumptions leading to the framework of flat FRW spacetimes and their evolution equations.

One option for this is modifying gravity to the point where the effects of such modifications influence cosmological evolution, breaking Assumption~\ref{a:relativity}. This approach has been studied for various modified gravity theories in~\cite{wangfinslerian,kazantzidisgravity,nunuesstructure,nilssonpreferredframe}.

A second way this can happen is when the evolution of the universe gains significant effects from deviations from perfect spatial flatness or isotropy. This would then break either or both of Assumptions~\ref{a:isotropy} and \ref{a:curvature}. Such an approach is studied for an "effective" curvature type contribution arising from matter clumping in~\cite{heinesensolving}. Approaches based on more explicit breaking of isotropy can be found in~\cite{amirhashchiinteracting}.

These ideas have in common that they, directly or indirectly, modify the relation between angular diameter distance and (some of the ingredients of) $H(z)$, putting the applicability of Equation~\ref{eq:distancerelation} into doubt.

A third option for breaking this would be an explicit global curvature. However, the authors are not aware of attempts to resolve the hubble tension this way, and while there seems to be some mild evidence for curvature in the CMB\cite{aghanimplanck, divalentinoplanck}, this in general is not supported by other datasets\cite{vagnozzieppur}.

\subsection{Breaking post-recombination assumptions}
This leaves three final assumptions we have not yet discussed: Breaking the relation between redshift and scale factor (Assumption~\ref{a:redshift}), having (dark) matter decay to some slower-diluting component, or having components in the universe whose density grows with redshift (both part of Assumption~\ref{a:growth}).

Both decaying (dark) matter solutions, as well as growing density-component solutions are very active areas of research. Because of this, we will not try to give an overview here of the candidates, but merely point to a number of recent papers~\cite{gomezupdate, brucksearching, valentinointeracting, beneventolate, divalentinodark, yangobservational, yangdynamical, choidegenerate} as a starting point for the interested reader, together with the more complete literature survey provided in~\cite{divalentinocosmology}. By breaking Assumption~\ref{a:growth}, such solutions can increase $H(0)$ relative to $H(z)$ (where $z > 0$) more than would otherwise be possible.

As to breaking the relation between redshift and scale factors, the authors are not aware of any proposed solutions exploring this avenue. As the resulting values of $H_0$ determined from the CMB are reasonably sensitive to these effects (due to the third-power dependency on the scale factor in $\Lambda$CDM) this might be an interesting area for further research. The $H_0$ shift in this situation arises because smaller than expected values for $a_{*}$ would give a larger range of integration in Equation~\ref{eq:distancerelation}, which in turn would allow $H(z)$ to be larger without $D_A$ dropping below the measured value.

\section{Conclusion}
We have shown that, for any theory to be able to solve the Hubble tension, it needs to break at least one of the Assumptions 1 through 7 defined above. This provides us with two main benefits. First of all, these conditions can be used as a tool to quickly determine whether a new proposed model has any hope of relieving the Hubble tension, and as a check on results for those models that break these conditions only in certain parts of their parameter space.

Furthermore, it allowed us to categorise the models already proposed. This showed us that breaking the standard relationship between redshift and scale factor can potentially solve the Hubble tension, but has so far not been explored in models. Exploration of mechanisms which can achieve such a changed link between redshift and scale factor could be an interesting avenue of further research.

One limitation of our work is that it constrains the theory space only from two very specific observations, the direct measurement of $H_0$ through distance ladders, and the interpretation of the CMB. Further research into general conditions on models resulting from other observations could provide useful tools for further exploring the space of models alleviating the Hubble tension. 

\section{Acknowledgements}
The authors would like to thank M. Postma and S. Vagnozzi for their feedback on early versions of this paper.

\bibliographystyle{unsrt}
\bibliography{conditions}
\end{document}